\documentclass[12pt]{article}
\usepackage{graphics}
\usepackage{graphicx}
\DeclareGraphicsExtensions{.pdf}
\usepackage{float} 
\usepackage{amsmath}
\usepackage{slashed}
\usepackage{amsfonts}   
\usepackage{amssymb}

\begin{document}
\date{}
\title{{\bf{\Large Nonrelativistic expansion of type IIA NS5 brane}}}
\author{
 {\bf {\normalsize Dibakar Roychowdhury}$
$\thanks{E-mail:  dibakarphys@gmail.com, dibakar.roychowdhury@ph.iitr.ac.in}}\\
 {\normalsize  Department of Physics, Indian Institute of Technology Roorkee,}\\
  {\normalsize Roorkee 247667, Uttarakhand, India}
\\[0.3cm]
}

\maketitle
\begin{abstract}
We carry out nonrelativistic expansion for NS5 brane based on a codimension two foliation of type IIA supergravity background. We simultaneously expand the world-volume fields in an appropriate $ 1/c^2 $ expansion together with the background fluxes. When put together, the resulting procedure yields a finite world-volume action for nonrelativistic NS5 brane that is coupled to String Newton-Cartan background.
\end{abstract}
\section{Overview and motivation}
Nonrelativistic (NR) limits of string theory \cite{Gomis:2000bd}-\cite{Gomis:2005pg} and D branes \cite{Gomis:2005bj} have witnessed a steady progress over the past couple of decades. In the literature, one finds two parallel approaches to NR strings. One of these approaches is based on taking a String Newton-Cartan (SNC) limit \cite{Bergshoeff:2018yvt}-\cite{Bergshoeff:2019pij} of relativistic strings over Lorentzian manifolds. The other approach is based on the null reduction procedure \cite{Harmark:2017rpg}-\cite{Harmark:2018cdl} of Lorentzian backgrounds\footnote{See \cite{Harmark:2019upf}-\cite{Hartong:2022dsx} for related works and \cite{Hartong:2022lsy}-\cite{Oling:2022fft} for recent reviews on the subject.}. 

SNC limit is characterised by a codimension two foliation of the Lorentzian target space \cite{Bergshoeff:2018yvt}. This implies that the tangent space is decomposed into two longitudinal directions labelled by the index $ A=0,1 $ and transverse directions labelled by $i=2, \cdots , 9$. We introduce set of translation generators $ H_A $ and $ P_i $ corresponding to the longitudinal and transverse spaces respectively. These generators act as an element of the string Galilei algebra that comprises in addition to the above generators (i) the string Galilean boost ($ G_{Ai} $), (ii) longitudinal Lorentz rotations ($ M_{AB} $) and (iii) transverse rotations ($ J_{ij} $). Gauging string Galilei algebra gives rise to a set of dynamical gauge fields $ \lbrace \tau_{\hat{\mu}}~^A,  m_{\hat{\mu}}~^A,  e_{\hat{\mu}}~^i \rbrace$ that characterise String Newton-Cartan (SNC) target space of the theory where $ \hat{\mu} $ is the ten dimensional spacetime index of type IIA. 

Here, $\tau_{\hat{\mu}}~^A $ is the longitudinal vielbein that introduces a (longitudinal) metric, $ \tau_{\hat{\mu}\hat{\nu}}=\tau_{\hat{\mu}}~^A\tau_{\hat{\nu}}~^B \eta_{AB} $. On the other hand, $ m_{\hat{\mu}}~^A $ is the gauge field associated with the non-central extension ($ Z_A $) of the string Galilean algebra \cite{Bergshoeff:2018yvt}. Finally, $ h^{\hat{\mu}\hat{\nu}}=e^{\hat{\mu}}~_i e^{\hat{\nu}}~_j\delta^{ij} $ is the co-metric associated with the transverse space where $ e^{\hat{\mu}}~_i $ is a projective inverse of $ e_{\hat{\mu}}~^i $.

The longitudinal directions of the tangent space possesses a global $ SO(1,1) $ Lorentz invariance which rotates the longitudinal axes ($ A, B =0,1$) under Lorentz transformation. On the other hand, the transverse space is invariant under a global $ SO(8) $ rotation that rotates transverse indices ($ i , j =2, \cdots , 9$). 

Recently, there has been a proposal to incorporate background NS-NS fluxes within the SNC framework of \cite{Bidussi:2021ujm}-\cite{Yan:2021lbe}. As the authors show, the presence of background fluxes modifies the algebra to F-string Galilei algebra \cite{Bidussi:2021ujm}. Down the line, NR limits \cite{Roychowdhury:2019qmp}-\cite{Bergshoeff:2022pzk}  of extended objects, those are naturally coupled to $ (p+1) $ forms, have also been explored in the literature. In \cite{Ebert:2021mfu}, the authors consider a codimension three foliation of $ D=11 $ supergravity target space while considering NR limits of membrane solutions of M theory. 

Following the above ideas, the purpose of the present paper is to workout the NR covariant world-volume action for a single NS5 brane \cite{Bandos:2000az}-\cite{Eyras:1998hn} using a codimension two foliation \cite{Bergshoeff:2018yvt} of type IIA supergravity target space. The NR NS5 brane naturally couples to the SNC target space that we mention above. Besides, it also couples to the background RR fluxes those are suitably expanded in a string $1/c^2$ expansion. On top of a codimension two foliation, we also expand the world-volume fields ($ X^{\hat{\mu }}$) of the relativistic theory using a string $ 1/c^2 $ expansion \cite{Hartong:2021ekg}. When put together, these two expansions result in a nice $ 1/c^2 $ expansion\footnote{This follows after a suitable rescaling of the NS5 brane tension.} where the leading order contribution appears at $\mathcal{O}(c^0)$. 

To start with, we choose a specific embedding for NS5 branes namely we set $ X^{\bar{a}}=\lbrace X^t, X^u \rbrace$ as the longitudinal axes and $ X^m $ being the remaining transverse directions of the NL target space\footnote{Notice that, this is a choice of embedding that fixes a particular orientation of the NS5 brane and thereby breaking the underlying diffeomerphism invariance of the theory. In other words, with this choice, two of the world-volume directions of the NS5 brane are extended along the two longitudinal axes of the non-Lorentzian target space while the remaining world-volume directions are along transverse axes.}. Later on, we generalise our calculations for arbitrary directions namely we set $\bar{a}=\hat{\mu}$ such that the longitudinal vielbeins can take values in any arbitrary direction. This results in a covariant NR world-volume action for NS5 brane. 

The organisation for the rest of the paper is as follows. In Section 2, we provide a brief introduction on the type IIA NS5 branes  \cite{Bandos:2000az}. In Section 3, we carry out NR expansion for the kinetic term of the NS5 brane Lagrangian. NR expansion for the WZ sector has been carried out in Section 4. We work out a covariant world-volume action and discuss its symmetries in Section 5. Finally, we conclude in Section 6.

\paragraph{Notation and convention.} Below we set the notation and other conventions for this paper. We denote $ \hat{\mu} $ as the usual ten dimensional spacetime index of type IIA and $ \hat{a}(=0, \cdots, 9) $ as flat indices equipped with a flat 10d metric $ \eta_{\hat{a}\hat{b}}$. Furthermore, we denote $ \sigma^a (a=0, \cdots , 5)$ as world-volume directions. The convention that we follow is that the time-time component of the world-volume metric is negative and the rest are positive.  As we choose a specific embedding for the NS5 brane, where we define $ \bar{a}=t,u $ as curved indices along two longitudinal directions. This introduces the following longitudinal metric, $ \tau_{\bar{a}\bar{b}}=\tau_{\bar{a}}~^A\tau_{\bar{b}}~^B \eta_{AB} $ with $ \eta_{AB}=\text{diag} (-1,1) $ and $ A=0,1 $ as longitudinal flat indices.
\section{Type IIA NS5 brane}
For the familiarity of the reader and completeness of the paper, we briefly discuss on M5 brane action and discuss its type IIA reduction \cite{Bandos:2000az} that leads to NS5 branes in $ D=10 $ supergravity. The M5 brane action in $ D=11 $ is given by\footnote{See \cite{Bandos:2000az} for details. } 
\begin{eqnarray}
 S_{M5}=T_{M5}\int d^6 \sigma \left[\sqrt{-\det (\hat{g}_{ab} + i H^{\ast}_{ab})} -\frac{\sqrt{-\det \hat{g}_{ab}}}{4 \partial_a \phi \partial^a \phi}\partial_c \phi H^{\ast cab}H_{abd}\partial^d \phi \right]+S_{WZ},
 \end{eqnarray}
where $ S_{WZ}  $ is the WZ term of the M5 brane \cite{Bandos:1997ui}-\cite{Bandos:2000az}
 \begin{eqnarray}
 \label{e2}
 S_{WZ} = T_{M5}\int \hat{C}^{(6)}+\frac{1}{2} F^{(3)}\wedge \hat{C}^{(3)}.
 \end{eqnarray}
Here, $ F^{(3)} = db^{(2)}$ is the world-volume three form flux while on the other hand, $  \hat{C}^{(3)} $ and $  \hat{C}^{(6)} $ correspond to the background three form and six form fluxes respectively.
 
The NS5 brane action is obtained following a type IIA reduction by compactifying one of the target space directions, $ X^{10}=\varrho $ where we assume that 10D fields do not depend on the compact direction. The dimensional reduction is defined by the following ansatz 
\begin{align}
\label{e5.1}
E^{\tilde{a}}= (E^{\hat{a}}, E^{10})~;~X^\mu = (X^{\hat{\mu}}, X^{10}=\varrho).
\end{align}

Here, we introduce the ten dimensional relativistic vielbein as
\begin{align}
\label{e5.2}
E^{\hat{a}} = e^{-\frac{\varphi}{3}}e_{\hat{\mu}}~^{\hat{a}}dX^{\hat{\mu}},
\end{align}
where the tenth component could be read out as
\begin{align}
E^{10} = e^{\frac{2\varphi}{3}}(d \varrho -\mathcal{A}_{\hat{\mu}} d X^{\hat{\mu}})=e^{\frac{2\varphi}{3}} \mathcal{F}.
\end{align}
Here, $ \varphi $ is the dilaton and $ \mathcal{A}_{\hat{\mu}} $ is the gauge field that lives in ten dimensions.

Using the vielbein \eqref{e5.2}, one can define $ D=10 $ metric as
\begin{align}
\label{e5.4}
g^{(10)}_{\hat{\mu}\hat{\nu}}=e_{\hat{\mu}}~^{\hat{a}}e_{\hat{\nu}}~^{\hat{b}}\eta_{\hat{a}\hat{b}}.
\end{align}

The six dimensional M5 brane world-volume metric is related to the six dimensional world-volume metric on the NS5 brane through the following relation \cite{Bandos:2000az}
\begin{align}
g^{(6)}_{ab}=e^{-\frac{2\varphi}{3}}(\tilde{g}_{ab}-e^{2\varphi}\mathcal{F}_a \mathcal{F}_b),
\end{align}
where $ g^{(6)}_{ab} $ is the M5 brane world-volume metric and $ \tilde{g}_{ab} $ is the NS5 brane world-volume metric and we use mostly -ve metric sign convention \cite{Bandos:2000az}.

Here, $ a,b=0, \cdots , 5 $ are the world-volume indices together with the six dimensional NS5 brane world-volume metric
\begin{align}
\label{e6}
&\tilde{g}_{ab}=g^{(10)}_{\hat{\mu}\hat{\nu}}\partial_a X^{\hat{\mu}}\partial_b X^{\hat{\nu}},\\
&\mathcal{F}_a=\partial_a \varrho - \mathcal{A}_{\hat{\mu}}\partial_a X^{\hat{\mu}}.
\end{align}

On the other hand, the inverse world-volume metrices are related as follows 
\begin{align}
g^{ab(6)}=e^{\frac{2\varphi}{3}}\left( \tilde{g}^{ab}+\frac{e^{2\varphi}\mathcal{F}^a \mathcal{F}^b}{1-e^{2\varphi}\mathcal{F}^2}\right). 
\end{align}

The self dual three form of the NS5 brane world-volume theory could be expressed as 
\begin{eqnarray}
\label{ee11}
H^{(3)}=db^{(2)}-C^{(3)}-B^{(2)}\wedge \mathcal{F},
\end{eqnarray}
where $ C^{(3)} $ and $ B^{(2)} $ are respectively the world-volume pull-backs of the background RR three form and NS-NS two form fluxes and $  \mathcal{F} =  \mathcal{F}_a d \sigma^a $ \cite{Bandos:2000az}.

Expanding the M5 brane Lagrangian upto quadratic order in the three form field strength ($ H_{abc} $), discarding the auxiliary fields ($ \phi $) and imposing the self-duality constraint \cite{Bandos:2000az}, one obtains the kinetic term for the NS5 brane world-volume theory as 
\begin{align}
\label{E5.11}
&S_{NS5}=-T_{NS5}\int d^6 \sigma e^{-2\varphi}\sqrt{-\det (\tilde{g}_{ab}-e^{2\varphi}\mathcal{F}_a \mathcal{F}_b)}\Big[ 1-\frac{1}{24}\Big( e^{2\varphi} H_{abc}H^{abc}\nonumber\\
&+\frac{3 e^{4\varphi}}{1-e^{2\varphi}\mathcal{F}^2}\mathcal{F}_a H^{abc}H_{bcd}\mathcal{F}^d\Big)+\cdots  \Big]+S_{WZ}\nonumber\\
&=- T_{NS5} \int d^6 \sigma \mathcal{L}^{(kin)}_{NS5}+S_{WZ},
\end{align}
where $ S_{WZ} $ is the Wess-Zumino (WZ) contribution that we elaborate in Section 4.
\section{The kinetic term}
We now discuss the NR expansion of NS5 brane based on the notion of a codimension two foliation \cite{Bergshoeff:2021bmc} of $ D=10 $ type IIA supergravity background. The expansion of the metric takes the following form
\begin{align}
g_{\hat{\mu}\hat{\nu}}^{(10)}=c^2\tau_{\hat{\mu}\hat{\nu}}(X)+h_{\hat{\mu}\hat{\nu}}(X)+\mathcal{O}(c^{-2}),
\end{align}
where $\eta_{AB}=\text{diag}(-1,1) $ is the two dimensional Lorentzian metric associated with the codimension two foliation.

We propose a $ 1/c^2 $ expansion \cite{Hartong:2021ekg} for the world-volume fields as
\begin{align}
\label{e5.55}
X^{\hat{\mu}}= x^{\hat{\mu}}+c^{-2}y^{\hat{\mu}}+\mathcal{O}(c^{-4}).
\end{align}

Before we proceed further, let us propose the following NS5 brane embedding. We choose two of the world-volume directions of the NR NS5 brane to be aligned along the longitudinal axes of the NL target space namely, $x^t= x^t (\sigma^0)$ and $x^u=x^u(\sigma^1)$. The remaining world-volume directions ($ \sigma^I $) are considered to be aligned along the transverse axes of the SNC background namely, $x^m = x^m (\sigma^I)$ where $I=2, \cdots ,5$. On the other hand, the $ 1/c^2 $ falloff could be of the generic form 
\begin{align}
y^{\hat{\mu}}=y^{\hat{\mu}}(\sigma^a).
\end{align}

Given the above facts and using \eqref{e5.55}, the $ D=10 $ background could be expressed as
\begin{align}
\label{e14}
&g_{tt}^{(10)}=c^2\tau_{tt}(x)+y^{\hat{\lambda}} \partial_{\hat{\lambda}} \tau_{tt}(x)+\mathcal{O}(c^{-2}),\\
&g_{uu}^{(10)}=c^2\tau_{uu}(x)+y^{\hat{\lambda}} \partial_{\hat{\lambda}} \tau_{uu}(x)+\mathcal{O}(c^{-2}),\\
&g_{mn}^{(10)}=h_{mn}(x)+\mathcal{O}(c^{-2}),
\label{e16}
\end{align}
where we assume the target space metric to be diagonal\footnote{Notice that, the metric $ h_{mn} $ is not always diagonal.}.

Using \eqref{e14}-\eqref{e16}, the pull back \eqref{e6} of the target space metric could be expressed as 
\begin{align}
\label{e18}
&\tilde{g}_{00}=c^2 \tau_{tt}(\partial_0 x^t)^2+ 2 \tau_{tt}\partial_0 x^t \partial_0 y^t + y^{\hat{\lambda}} \partial_{\hat{\lambda}} \tau_{tt}(\partial_0 x^t)^2 + \mathcal{O}(c^{-2}),\\
\label{e19}
&\tilde{g}_{11}=c^2 \tau_{uu}(\partial_1 x^u)^2+ 2 \tau_{uu}\partial_1 x^u \partial_1 y^u + y^{\hat{\lambda}} \partial_{\hat{\lambda}} \tau_{uu}(\partial_1 x^u)^2 + \mathcal{O}(c^{-2}),\\
&\tilde{g}_{IJ}=h_{mn}\partial_I x^m \partial_J x^n + \mathcal{O}(c^{-2}).
\label{e20}
\end{align}

Finally, the $ U(1) $ gauge field and the embedding scalar are expanded as
\begin{align}
&\mathcal{A}_{\hat{\mu}}= \mathcal{A}_{\hat{\mu}}(x)+c^{-2}y^{\hat{\lambda}} \partial_{\hat{\lambda}} \mathcal{A}_{\hat{\mu}}(x)+\mathcal{O}(c^{-4}),\\
&\varrho =  \varrho(x)+c^{-2}y^{\hat{\lambda}} \partial_{\hat{\lambda}} \varrho(x)+\mathcal{O}(c^{-4}).
\end{align}

In what follows, we choose to work with the following ansatz for the target space fields
\begin{align}
\varrho = \varrho (x^{\bar{m}})~;~\mathcal{A}_{\hat{\mu}}=(a_t (x^{\bar{m}}),a_u(x^{\bar{m}}),0,\cdots ,0)~;~x^{\bar{m}}=x^t, x^u
\end{align}
which results into an expansion of the following form ($ a=0,1 $)
\begin{align}
&\mathcal{F}_a =f_a (x^{\bar{m}})+c^{-2}\Lambda_{a}(x^{\bar{m}}, y^{\bar{m}}) +\mathcal{O}(c^{-4}).
\end{align}

We denote the above functions as
\begin{align}
&f_0 (x)=\partial_0 \varrho - a_t \partial_0 x^t ~;~ f_1 (x^u)=\partial_1 \varrho - a_u \partial_1 x^u\\
&\Lambda_0 (x, y)=\partial_0 y^{\bar{m}} \partial_{\bar{m}} \varrho + y^{\bar{m}} \partial_0 \partial_{\bar{m}} \varrho -a_{\bar{m}}\partial_0 y^{\bar{m}} -y^{\bar{m}} \partial_{\bar{m}} a_t \partial_0 x^t,\\
&\Lambda_1 (x, y)=\partial_1 y^{\bar{m}} \partial_{\bar{m}} \varrho + y^{\bar{m}} \partial_1 \partial_{\bar{m}} \varrho -a_{\bar{m}}\partial_1 y^{\bar{m}} -y^{\bar{m}} \partial_{\bar{m}} a_u \partial_1 x^u.
\end{align}

On the other hand, for the remaining world-volume directions, one finds
\begin{align}
\mathcal{F}_I = c^{-2}(\partial_I y^{\bar{m}}\partial_{\bar{m}}\varrho- a_{\bar{m}}\partial_I y^{\bar{m}}).
\end{align}

In order to estimate the world-volume determinant in the NR limit of the NS5 brane, it is always useful to calculate the following entities those comprise the determinant
\begin{align}
\label{e29}
&\tilde{g}_{00} - e^{2\varphi}\mathcal{F}^2_0 = c^2 \Big[ \tau_{tt}(\partial_0 x^t)^2 - e^{2 \tilde{\varphi}}f_0^2  \Big]+\Delta^{(0)}(x,y)+\mathcal{O}(c^{-2}),\\
&\tilde{g}_{11} - e^{2\varphi}\mathcal{F}^2_1 = c^2 \Big[ \tau_{uu}(\partial_1 x^u)^2 - e^{2 \tilde{\varphi}}f_1^2  \Big]+\Delta^{(1)}(x,y)+\mathcal{O}(c^{-2}),\\
&\tilde{g}_{IJ} - e^{2\varphi}\mathcal{F}_I  \mathcal{F}_J =h_{IJ}-e^{2 \tilde{\varphi}}(\partial_I y^{\bar{m}}\partial_{\bar{m}}\varrho- a_{\bar{m}}\partial_I y^{\bar{m}})\nonumber\\&(\partial_J y^{\bar{n}}\partial_{\bar{n}}\varrho- a_{\bar{n}}\partial_J y^{\bar{n}})+\mathcal{O}(c^{-2})\equiv  h_{IJ}-e^{2\tilde{\varphi}}\Gamma_{IJ}+\mathcal{O}(c^{-2}).
\label{e31}
\end{align}

Here, we choose the NR expansion \cite{Bergshoeff:2021bmc} for the dilaton to be $\varphi = \tilde{\varphi}+ \log c$ and denote $ h_{IJ}= h_{mn}(x)\partial_I x^m \partial_J x^n$. We also denote the above functions as
\begin{align}
\Delta^{(a)}=2 \tau_{\bar{m}\bar{n}}\partial_a x^{\bar{m}}\partial_a y^{\bar{n}}+y^{\hat{\lambda}}\partial_{\hat{\lambda}}\tau_{\bar{m}\bar{n}}\partial_a x^{\bar{m}}\partial_a x^{\bar{n}}-2 e^{2 \tilde{\varphi}}f_a \Lambda_a,
\end{align}
where $ a $ is not summed over and could take values either $ 0 $ or $ 1 $.

Following the steps as alluded to the above \eqref{e29}-\eqref{e31}, it is straightforward to calculate
\begin{align}
\label{e33}
&\sqrt{-\det (\tilde{g}_{ab}-e^{2\varphi}\mathcal{F}_a \mathcal{F}_b)}\nonumber\\
&=c^2 \sqrt{- \tau_{00}\tau_{11}+e^{2\tilde{\varphi}}\Big( \tau_{00}f^2_1 + \tau_{11}f^2_0\Big)}\sqrt{\det  (h_{IJ}-e^{2\tilde{\varphi}}\Gamma_{IJ})}+\mathcal{O}(c^0),
\end{align}
where we denote $ \tau_{00}=\tau_{tt}(\partial_0 x^t)^2 $ and so on.

The world-volume three form fluxes, on the other hand, can be expanded as
\begin{align}
H_{abc}=\hat{H}^{(3)}_{abc}(x)+\mathcal{O}(c^{-2}).
\end{align}

In order to raise or lower the world-volume indices, the first step would be to introduce the inverse metric on the world-volume. For longitudinal world-volume axes, the metric (\eqref{e18} and \eqref{e19}) and its inverse could be schematically expressed as
\begin{align}
&\tilde{g}_{ab}=c^2 \tau_{ab}+\varpi_{ab}+\cdots,\\
&\tilde{g}^{ab}=\varpi^{ab}+c^{-2}\tau^{ab}+\cdots,
\end{align}
such that the following identities should hold in order to preserve the orthonormality
\begin{align}
\tau_{ab}\varpi^{bc}=0~;~\tau_{ab}\tau^{bc}+\varpi_{ab}\varpi^{bc}=\delta_a^c.
\end{align}

For the transverse world-volume axes, the inverse metric of \eqref{e20} looks trivial
\begin{align}
\tilde{g}^{IJ}=h^{IJ}.
\end{align}

This leads to the following set of expansions for the NS sector
\begin{align}
\label{e39}
&e^{2\varphi}H_{abc}H^{abc}= c^2 e^{2 \tilde{\varphi}}\varpi^{am}\varpi^{bn}\varpi^{cl}\hat{H}_{abc}\hat{H}_{mnl}+\mathcal{O}(c^0),\\
&\frac{3 e^{4\varphi}}{1-e^{2\varphi}\mathcal{F}^2}\mathcal{F}_a H^{abc}H_{bcd}\mathcal{F}^d=-\frac{3 c^2}{f^2} e^{2 \tilde{\varphi}}\varpi^{am}\varpi^{bn}\varpi^{cl}\varpi^{dk}f_a\hat{H}_{bcd}\hat{H}_{mnl}f_k +\mathcal{O}(c^0),
\label{e40}
\end{align}
where we denote, $ f^2 = \varpi^{ab}f_a f_b+\mathcal{O}(c^{-2}) $.

Using \eqref{e33}, \eqref{e39} and \eqref{e40}, the NR world-volume action corresponding to the kinetic term could be schematically expanded as
\begin{align}
S^{(kin)}_{NR}=T_{NR}\int d^6\sigma \Big(L^{{(kin)}}_{LO}+c^{-2}L^{{(kin)}}_{NLO}+\cdots  \Big),
\end{align}
where for the rest of the analysis we retain ourselves only to the LO Lagrangian density. Here, we define the NS5 brane tension in the NR limit as $T_{NR}=c^2 T_{NS5}  $.

The LO Lagrangian density could be expressed as
\begin{align}
\label{e42}
&L^{{(kin)}}_{LO}=\frac{1}{24}\sqrt{- \tau_{00}\tau_{11}+e^{2\tilde{\varphi}}\Big( \tau_{00}f^2_1 + \tau_{11}f^2_0\Big)}\sqrt{\det  (h_{IJ}-e^{2\tilde{\varphi}}\Gamma_{IJ})}\nonumber\\
&\times \varpi^{am}\varpi^{bn}\varpi^{cl}\Big(\hat{H}_{abc}\hat{H}_{mnl}-\frac{3}{f^2} \varpi^{dk}f_a\hat{H}_{bcd}\hat{H}_{mnl}f_k \Big).
\end{align}
\section{WZ sector} 
The WZ action for relativistic NS5 brane has its form
\begin{align}
S_{WZ}=T_{NS5}\int  \mathcal{L}_{WZ},
\end{align}
where the WZ Lagrangian is expressed as 
\begin{align}
\label{ee46}
\mathcal{L}_{WZ}=B^{(6)}+C^{(5)}\wedge \mathcal{F}+\frac{1}{2}db^{(2)}\wedge C^{(3)}+\frac{1}{2}db^{(2)}\wedge B^{(2)}\wedge \mathcal{F},
\end{align}
where $ b^{(2)} $ is the world-volume two form \cite{Bandos:2000az}. Here, the six form $ B^{(6)} $ and the five form $ C^{(5)} $ fluxes in $ D=10 $ are produced following the dimension reduction of the six form flux in M theory. On the other hand, $ C^{(3)} $ and $ B^{(2)} $ are produced following the dimension reduction of the three form flux of M theory background \cite{Bandos:2000az}.

In order define a consistent $ 1/c^2 $ expansion, we propose the following NR scaling\footnote{The $ c^2 $ scaling stems from the fact that $\ell^{(2)}  $ is expressed in terms of two longitudinal vielbeins ($ \tau_{\hat{\mu}}~^A $) each of which comes with a factor of $ c $ as an artefact of the SNC expansion.}
\begin{align}
\label{e5.42}
&\mathcal{F}= \tilde{\mathcal{F}}(x)+\mathcal{O}(c^{-2})~;~\tilde{\mathcal{F}}(x)= f_a (x^{\bar{a}})d\sigma^a,\\
\label{ee47}
&B^{(2)}=c^2 \ell^{(2)}(x)+\hat{B}^{(2)}(x)+y^{\hat{\lambda}}\partial_{\hat{\lambda}}\ell^{(2)}(x)+\mathcal{O}(c^{-2}),\\
&db^{(2)}=d\hat{b}^{(2)}(x)+\mathcal{O}(c^{-2}),\\
\label{ee49}
&C^{(3)}=-c^2 \ell^{(2)}(x)\wedge  \tilde{\mathcal{F}}(x)+\hat{C}^{(3)}(x)-\Sigma^{(3)}(x,y)+\mathcal{O}(c^{-2}),\\
&C^{(5)}=-c^2 \hat{C}^{(3)}(x)\wedge \ell^{(2)}(x)+ \hat{C}^{(5)}(x)-\Sigma^{(5)}(x,y)+\mathcal{O}(c^{-2}),\\
&B^{(6)}=-c^2 d\hat{b}^{(2)}(x) \wedge \ell^{(2)}(x)\wedge \tilde{\mathcal{F}}(x)+\hat{B}^{(6)}(x)+\mathcal{O}(c^{-2}),
\label{e5.47}
\end{align}
where for example, $\Sigma^{(3)}(x,y)=y^{\hat{\lambda}}\partial_{\hat{\lambda}}(\ell^{(2)}(x)\wedge  \tilde{\mathcal{F}}(x))$ denotes derivative corrections that contribute at $\mathcal{O}(c^{0})$. Similar remarks hold for other $\Sigma^{(n)}(x,y)$ terms in the expansion.

The above expansions \eqref{e5.42}-\eqref{e5.47} are defined in terms of a longitudinal two form
\begin{align}
&\ell^{(2)} = \frac{1}{2!}\tau_{\bar{m}}~^A (x)\tau_{\bar{n}}~^B (x)\varepsilon_{AB}d x^{\bar{m}}\wedge d x^{\bar{n}}+\mathcal{O}(c^{-2})\nonumber\\
&=\frac{1}{2!} \ell_{\bar{m}\bar{n}}\varepsilon^{ab}\partial_a x^{\bar{m}}\partial_b x^{\bar{n}}+\mathcal{O}(c^{-2})\nonumber\\
&= \ell^{(2)} (x)+\mathcal{O}(c^{-2}).
\end{align}

Combining the above expansions \eqref{e5.42}-\eqref{e5.47}, the LO WZ contribution turns out to be
\begin{align}
\label{ee54}
&L^{(wz)}_{LO}=d\hat{b}^{(2)}(x)\wedge \tilde{\mathcal{F}} (x)\wedge \ell^{(2)}(x)-\hat{C}^{(3)}(x)\wedge \ell^{(2)}(x)\wedge  \tilde{\mathcal{F}} (x).
\end{align}
\section{Covariant formulation and its symmetries}
We now provide a covariant formulation of NR NS5 brane action where we do not restrict longitudinal vielbeins to a particular direction. To begin with, we expand the ten dimensional metric (\ref{e5.4}) as well as the background fluxes following the prescription of \cite{Ebert:2021mfu}
\begin{align}
\label{EE5.13}
g_{\hat{\mu}\hat{\nu}}^{(10)}=c^2 \tau_{\hat{\mu}\hat{\nu}}+h_{\hat{\mu}\hat{\nu}}~;~\tau_{\hat{\mu} \hat{\nu}}= \tau_{\hat{\mu}}~^A\tau_{\hat{\nu}}~^B \eta_{AB},
\end{align}
where $ \tau_{\hat{\mu}}~^A ~(A=0, 1)$ are the longitudinal vielbeins of ten dimensional target space.

Like before, we also expand the world-volume fields as \cite{Harmark:2019upf}, \cite{Hartong:2021ekg}
\begin{align}
\label{EE5.16}
X^{\hat{\mu}}= x^{\hat{\mu}}+c^{-2}y^{\hat{\mu}}+\mathcal{O}(c^{-4}).
\end{align}

The above expansion (\ref{EE5.16}) further modifies (\ref{EE5.13}) as
\begin{align}
\label{EE5.17}
g_{\hat{\mu}\hat{\nu}}^{(10)}=c^2\tau_{\hat{\mu}\hat{\nu}}(x)+H_{\hat{\mu}\hat{\nu}}(x, y)+ \cdots ,
\end{align}
where we ignore the \emph{second} order fluctuations\footnote{This assumption is valid under the circumstances in which the String Newton-Cartan (SNC) fields are considered to be \emph{slowly} varying functions of the target space coordinates.} of $\tau_{\hat{\mu}}~^A$ to yield
\begin{align}
\label{ee5.19}
H_{\hat{\mu}\hat{\nu}}(x , y)=h_{\hat{\mu}\hat{\nu}}(x)+y^\lambda \partial_\lambda \tau_{\hat{\mu}\hat{\nu}}(x).
\end{align}

Using (\ref{EE5.17}), the pull back of the target space metric could be expressed as
\begin{eqnarray}
\label{e5.10}
\tilde{g}_{ab}=c^2 \bar{\tau}_{ab}^{(2)}+\bar{\Sigma}_{ab}^{(0)}+\mathcal{O}(c^{-2}) ,
\end{eqnarray}
where bars take care of the fact that these are the world-volume fields in a $ 1/c $ expansion of NS5 branes that are embedded into a ten dimensional SNC background
\begin{align}
\label{5.21}
&\bar{\tau}_{ab}^{(2)} = \tau_{\hat{\mu} \hat{\nu}}(x)\partial_a x^{\hat{\mu}}\partial_b x^{\hat{\nu}},\\
&\bar{\Sigma}_{ab}^{(0)}=\tau_{\hat{\mu} \hat{\nu}}(x)(\partial_ a x^{\hat{\mu}}\partial_by^{\hat{\nu}}+\partial_ a y^{\hat{\mu}}\partial_b x^{\hat{\nu}})+H_{\hat{\mu}\hat{\nu}}(x , y)\partial_a x^{\hat{\mu}}\partial_b x^{\hat{\nu}}.
\label{5.22}
\end{align}

Next, we work out the NR expansion of $ \mathcal{F}_a $. This is obtained by introducing a $ 1/c $ expansion for the background $ U(1) $ gauge fields ($ \mathcal{A}_{\hat{\mu}} $) and the world-volume scalar ($ \varrho $) 
\begin{align}
\label{e5.17}
&\mathcal{A}_{\hat{\mu}}= \mathcal{A}_{\hat{\mu}}(x)+c^{-2}y^{\hat{\lambda}} \partial_{\hat{\lambda}} \mathcal{A}_{\hat{\mu}}(x)+\mathcal{O}(c^{-4}),\\
&\varrho =  \varrho(x)+c^{-2}y^{\hat{\lambda}} \partial_{\hat{\lambda}} \varrho(x)+\mathcal{O}(c^{-4}).
\label{e5.18}
\end{align}

Using (\ref{e5.17})-(\ref{e5.18}), we find out a $ 1/c $ expansion of the following form
\begin{align}
\label{e5.19}
&e^{2\varphi}\mathcal{F}_a \mathcal{F}_b= c^2 e^{2\tilde{\varphi}(x)}\Big[f_a^{(0)}f_b^{(0)}+c^{-2}\left(f_a^{(0)}f_b^{(-1)}+f_a^{(-1)}f_b^{(0)} \right)+\cdots\Big],
\end{align}
where we choose the NR expansion for the dilaton to be \cite{Bergshoeff:2021bmc}, \cite{Ebert:2021mfu}
\begin{align}
\varphi = \tilde{\varphi}+ \log c.
\end{align}

The above functions are identified as 
\begin{align}
\label{5.28}
&f_a^{(0)}(x)=\partial_a \varrho -\mathcal{A}_{\hat{\mu}}(x)\partial_a x^{\hat{\mu}},\\
&f_a^{(-1)}(x,y)=y^{\hat{\lambda}}\partial_{\hat{\lambda}} \varrho -y^{\hat{\lambda}} \partial_{\hat{\lambda}} \mathcal{A}_{\hat{\mu}}(x)\partial_a x^{\hat{\mu}}-\mathcal{A}_{\hat{\mu}}(x)\partial_a y^{\hat{\mu}}.
\label{5.29}
\end{align}

Combining (\ref{e5.10}) and (\ref{e5.19}), we finally obtain\footnote{Notice that, \eqref{ee68} boils down to \eqref{e29}-\eqref{e31} when we switch to a particular embedding.} 
\begin{align}
\label{ee68}
&\tilde{g}_{ab}-e^{2\varphi}\mathcal{F}_a \mathcal{F}_b =c^2 \Big [ \left( \bar{\tau}_{ab}^{(2)} -e^{2\tilde{\varphi}}f_a^{(0)}f_b^{(0)}\right) \nonumber\\
&+c^{-2}\left(\bar{\Sigma}^{(0)}_{ab}-e^{2\tilde{\varphi}}(f_a^{(0)}f_b^{(-1)}+f_a^{(-1)}f_b^{(0)}) \right) \Big]+\mathcal{O}(c^{-2}).
\end{align}

Finally, the determinant in (\ref{e33}) takes the following form 
\begin{align}
&\sqrt{-\det (\tilde{g}_{ab}-e^{2\varphi}\mathcal{F}_a \mathcal{F}_b)}\nonumber\\
&=c^2\sqrt{-\det \left( \bar{\tau}_{ab}^{(2)} -e^{2\tilde{\varphi}}f_a^{(0)}f_b^{(0)}\right)}\Big[ 1\nonumber\\
&+\frac{c^{-2} }{2}\tilde{\mathcal{N}}^{ab}\left(\bar{\Sigma}^{(0)}_{ab}-e^{2\tilde{\varphi}}(f_a^{(0)}f_b^{(-1)}+f_a^{(-1)}f_b^{(0)}) \right) +\cdots\Big]\nonumber\\& \equiv c^2 \sqrt{- \det \tilde{\Delta}_{ab}}\left[1 +\frac{c^{-2} }{2}\tilde{\mathcal{N}}^{ab} \Pi_{ab} + \cdots\right],
\end{align}
where $ \tilde{\mathcal{N}}^{ab}$ are the matrices inverse to $\tilde{\Delta}_{ab}= \bar{\tau}_{ab}^{(2)} -e^{2\tilde{\varphi}}f_a^{(0)}f_b^{(0)}$. 

Next, we expand all the remaining terms in (\ref{E5.11}) which eventually which reveals identical expansions \eqref{e39}-\eqref{e40} as before. Combining all the above pieces together and following a suitable rescaling of the brane tension, we find an expansion of the following form
\begin{align}
S^{(kin)}_{NR}=T_{NR}\int d^6\sigma \Big(L^{{(kin)}}_{LO}+c^{-2}L^{{(kin)}}_{NLO}+\cdots  \Big),
\end{align}
where we denote NR kinetic Lagrangian density at LO as
\begin{align}
\label{e5.39}
&L^{{(kin)}}_{LO}=\frac{1}{24} \sqrt{- \det \tilde{\Delta}_{ab}}\varpi^{am}\varpi^{bn}\varpi^{cl}\Big(\hat{H}_{abc}\hat{H}_{mnl}-\frac{3}{f^2} \varpi^{dk}f_a\hat{H}_{bcd}\hat{H}_{mnl}f_k \Big).
\end{align}
Finally, the LO WZ contribution \eqref{ee54} appears as before, except for the fact that the longitudinal two form is now expressed as, $ \ell^{(2)}\sim \ell_{\hat{\mu}\hat{\nu}}\varepsilon^{ab}\partial_a x^{\hat{\mu}}\partial_b x^{\hat{\nu}} $. 

\paragraph{Symmetries.} We restrict ourselves to the strict NR limit and discuss symmetries of the LO theory \eqref{e5.39} and \eqref{ee54}. The LO theory is manifestly invariant under world-volume and ten dimensional target space diffeomorphisms. It is also noteworthy to mention that, $ \delta \bar{\tau}_{ab}^{(2)}=0 $ under $ SO(1,1) $ that rotates (longitudinal) tangent space indices $ A=0,1 $. Following the orthonormality condition $  \bar{\tau}_{ab}^{(2)}\varpi^{bc}=0 $, one could further show that $ \varpi^{ab} $ is also invariant and hence the LO theory is invariant under $ SO(1,1) $. However, to show that the LO theory possesses NR invariance, one must explore its properties under string Galilei boost \cite{Bergshoeff:2018yvt}, \cite{Bidussi:2021ujm} and in particular show its invariance under the above transformation.

Next, we introduce string Galilei boost transformations \cite{Bergshoeff:2018yvt} $\delta_\Sigma  \tau_{\hat{\mu}}~^A =0$. This leaves the longitudinal metric invariant namely, $\delta_\Sigma  \tau_{\hat{\mu} \hat{\nu}} =0$ and thereby preserves the world-volume pull-back $\delta_\Sigma   \bar{\tau}_{ab}^{(2)}=0$. This altogether preserves $\delta_\Sigma \tilde{\Delta}_{ab}=0$, assuming that the variation of the remaining fields ($ f^{(0)}_a $ and $ f^{(-1)}_a $) are zero under Galilean boost. Following the orthonormality condition $  \bar{\tau}_{ab}^{(2)}\varpi^{bc}=0 $, one could further argue that $ \delta_\Sigma\varpi^{ab} =0$.

Clearly, the invariance of the kinetic term \eqref{e5.39} is solely determined by the variation of the world-volume three form flux ($ \hat{H}_{abc} $) under the string Glalilei boost. Using \eqref{ee11}, \eqref{ee47} and \eqref{ee49}, one finds the self dual three form flux
\begin{align}
&H^{(3)}=d\hat{b}^{(2)}-\hat{C}^{(3)}+y^{\hat{\lambda}}\ell^{(2)}\wedge  \partial_{\hat{\lambda}}\tilde{\mathcal{F}}-\hat{B}^{(2)}\wedge \tilde{\mathcal{F}}+ \mathcal{O}(c^{-2})\nonumber\\
&= \hat{H}_{abc} d\sigma^a \wedge d\sigma^b \wedge d\sigma^c +\mathcal{O}(c^{-2}).
\end{align}

Under Galilean boost, $ \delta_\Sigma \ell_{\hat{\mu}\hat{\nu}}=\delta_\Sigma(\tau_{\hat{\mu}}~^A \tau_{\hat{\nu}}~^B \varepsilon_{AB})=0 $. On the other hand, considering the variation of the kinetic term \eqref{e5.39} to be zero under the string Galilei boost, we are finally left with the following variation condition for the world-volume three form flux 
\begin{align}
\delta_\Sigma H^{(3)}=\delta_\Sigma \hat{C}^{(3)}+\delta_\Sigma \hat{B}^{(2)}\wedge \tilde{\mathcal{F}}-\delta_\Sigma d \hat{b}^{(2)}=0.
\end{align}

On the other hand, the invariance of the LO WZ contribution \eqref{ee54} yields
\begin{align}
\label{ee73}
\delta_\Sigma d \hat{b}^{(2)}=-\delta_\Sigma \hat{C}^{(3)}.
\end{align}

Using \eqref{ee73}, one finally obtains
\begin{align}
\delta_\Sigma \hat{C}^{(3)}+\frac{1}{2}\delta_\Sigma \hat{B}^{(2)}\wedge \tilde{\mathcal{F}}=0=\frac{1}{2}\delta_\Sigma H^{(3)}.
\end{align}

Splitting into components, and setting $ \delta_\Sigma \hat{H}_{abc}=0 $, finally one fixes the transformation of the RR three form under string Galilei boost
\begin{align}
\label{ee74}
\delta_\Sigma \hat{C}^{(3)}_{abc}=-\frac{1}{2 . 3!}\Big(\delta_\Sigma \hat{B}^{(2)}_{ab}\tilde{\mathcal{F}}_{c}-\delta_\Sigma \hat{B}^{(2)}_{bc}\tilde{\mathcal{F}}_{a} +\delta_\Sigma \hat{B}^{(2)}_{ca}\tilde{\mathcal{F}}_{b} \Big).
\end{align}

One could further simplify the R.H.S. of \eqref{ee74} using the transformation of NS-NS two form under Galilei boost. A close comparison between the expansion \eqref{ee47} and the background NS-NS field of \cite{Bidussi:2021ujm} reveals that $ \hat{B}^{(2)}_{\hat{\mu}\hat{\nu}}  $ is precisely the two form $ m_{\hat{\mu} \hat{\nu}} $ that is expressed in terms of the TSNC fields namely, $  \hat{B}^{(2)}_{\hat{\mu}\hat{\nu}} =m_{\hat{\mu} \hat{\nu}} $ such that
\begin{align}
m_{\hat{\mu} \hat{\nu}}=\eta_{AB}\tau_{[\hat{\mu}}~^A \pi_{\hat{\nu}]}~^B + \delta_{ij}e_{[\hat{\mu}}~^i \pi_{\hat{\nu}]}~^j,
\end{align}
where $ \pi_{\hat{\mu}}~^i $ are the gauge fields associated to F-string Galilei algebra \cite{Bidussi:2021ujm}.

Considering the transformation under Galilei boost \cite{Bidussi:2021ujm}
\begin{align}
\delta_\Sigma\hat{B}^{(2)}_{\hat{\mu}\hat{\nu}} =\delta_\Sigma m_{\hat{\mu} \hat{\nu}} = -2\varepsilon_{AB}\Sigma^B~_i \tau_{[\hat{\mu}}~^A e_{\hat{\nu}]}~^i,
\end{align}
one finds the following transformation for the world-volume pull-back of the NS two form
\begin{align}
\label{ee77}
\delta_\Sigma\hat{B}^{(2)}_{ab}=-2\varepsilon_{AB}\Sigma^B~_i \tau_{[a}~^A e_{b]}~^i,
\end{align}
where $ \tau_{a}~^A =\tau_{\hat{\mu}}~^A \partial_a x^{\hat{\mu}}$ and so on.

Using \eqref{ee77}, one could finally write down the transformation of RR three form on the world-volume such that the LO Lagrangian is invariant under string Galilei boost 
\begin{align}
\label{e79}
\delta_\Sigma \hat{C}^{(3)}_{abc}=-\frac{1}{3!}\varepsilon_{AB}\Sigma^B~_i \Big( \tau_{[a}~^A e_{b]}~^i\tilde{\mathcal{F}}_{c}- \tau_{[b}~^A e_{c]}~^i\tilde{\mathcal{F}}_{a} + \tau_{[c}~^A e_{a]}~^i\tilde{\mathcal{F}}_{b} \Big).
\end{align}

\paragraph{Note added.} Proving the invariance of the NR action, the other way around, looks quite non-trivial. For example, a naive variation of the kinetic Lagrangian yields
\begin{align}
\delta_\Sigma L^{{(kin)}}_{LO}=\frac{1}{12} \sqrt{- \det \tilde{\Delta}_{ab}}\varpi^{am}\varpi^{bn}\varpi^{cl}(\delta_\Sigma\hat{H}_{abc})\hat{H}_{mnl}+\cdots,
\end{align}
where we denote the variation under Galilean boost,
\begin{align}
\delta_\Sigma\hat{H}_{abc}=\partial_{[a}\delta_\Sigma \hat{b}_{bc]}-\delta_\Sigma \hat{C}^{(3)}_{abc}-\delta_\Sigma \hat{B}^{(2)}_{[ab}\tilde{\mathcal{F}}_{c]}.
\end{align}

On the other hand, the variation of the WZ term yields
\begin{align}
&\delta_\Sigma L^{(wz)}_{LO}=\delta_\Sigma d\hat{b}^{(2)}(x)\wedge \tilde{\mathcal{F}} (x)\wedge \ell^{(2)}(x)-\delta_\Sigma \hat{C}^{(3)}(x)\wedge \ell^{(2)}(x)\wedge  \tilde{\mathcal{F}} (x)\nonumber\\
&=\partial_{[a}\delta_\Sigma b^{(2)}_{bc}\tilde{\mathcal{F}}_d \ell^{(2)}_{ef]}d\sigma^a \wedge \cdots \wedge d\sigma^f +\delta_\Sigma \hat{C}^{(3)}_{[abc}\tilde{\mathcal{F}}_d\ell^{(2)}_{ef]}d\sigma^a \wedge \cdots \wedge d\sigma^f.
\end{align}

Given the above expressions, it does not seem plausible to show the invariance of the NR action as a result of the mutual cancellation between the kinetic and WZ contributions which should take place by virtue of using the expansions of the background fields. We therefore find it to be more convenient to claim the individual variations to be zero under string Galilei boost and derive the transformation rule for 10d RR flux from there.

The above formula \eqref{e79}, could however be motivated starting from an expansion in the eleven dimensional M-theory background \cite{Blair:2021waq}
\begin{align}
\label{e83}
\delta_\Sigma C^{(3)}_{\mu \nu \lambda}\Big |_{11d} \sim -\epsilon_{ABC}\Sigma_{[\mu}~^A \tau_\nu~^B \tau_{\lambda]}~^C,
\end{align}
and thereby following a dimensional reduction along a particular axis, $ X^{10}=\varrho $.

Using \eqref{e83}, it is trivial to show
\begin{align}
\label{e84}
\delta_\Sigma C^{(3)}_{abc}\Big |_{11d} \sim -\epsilon_{ABC}\Sigma^A~_i e_{[\mu}~^i \tau_\nu~^B \tau_{\lambda]}~^C \partial_a X^\mu \partial_b X^\nu \partial_c X^\lambda,
\end{align}
where we use, $ \Sigma_\mu~^A = \Sigma^A~_i e_\mu~^i $. 

Following a dimensional reduction, the above expression \eqref{e84} boils down into a structure similar to that of  \eqref{e79} (along with a bunch of other terms as a consequence of reducing from a higher dimensional theory) where we identify
\begin{align}
\tilde{\mathcal{F}}_a \sim \tau_{X^{10}}^{10}\partial_a\varrho.
\end{align}
\section{Summary and final remarks}
In the present paper, we construct a NR world-volume action for NS5 branes considering String Newton-Cartan (SNC) limit \cite{Bergshoeff:2018yvt} of type IIA supergravity. To start with, we choose a specific embedding for NS5 branes which was generalised later on. These NR branes are coupled to String Newton-Cartan (SNC) backgrounds that are characterized by a codimension two foliation of the target space. 

The above NR reduction, need not have to be the unique procedure and could also be tested from different perspectives, for example, the \emph{null} reduction approach of \cite{Harmark:2017rpg}-\cite{Harmark:2018cdl}. It would be nice to see a mutual compatibility \cite{Harmark:2019upf} between these two limits. Other interesting directions would be to explore T-duality properties of the LO action along with its super-symmetric extensions. We leave all these issues for future endeavours.
\paragraph{Acknowledgments.}
The author is indebted to the authorities of IIT Roorkee for their unconditional support towards researches in basic sciences. The author acknowledges The Royal Society, UK for financial assistance.

\end{document}